\documentclass[twocolumn,showpacs,superscriptaddress,prb,notitlepage]{revtex4-1}

\usepackage{graphicx}
\usepackage{color}
\usepackage{amsmath,amsthm,amssymb}
\usepackage{braket}
\usepackage[colorlinks,citecolor=blue,linkcolor=red]{hyperref}
\usepackage{multirow}
\usepackage{ulem}
\usepackage{cancel}

\begin{document}

\title{Competing Magnetic Ground States in Copper-Doped Pb\textsubscript{10}P\textsubscript{6}O\textsubscript{25}}

\author{Lin Hou}
\affiliation{Theoretical Division, Los Alamos National Laboratory, Los Alamos, New Mexico 87545, USA}
\affiliation{Department of Physics and Engineering Physics, Tulane University, New Orleans, 70118, Louisiana, USA}

\author{Kevin Allen}
\affiliation{Theoretical Division, Los Alamos National Laboratory, Los Alamos, New Mexico 87545, USA}
\affiliation{Department of Physics and Astronomy, Rice University, Houston, 77005, Texas, USA}

\author{Christopher Lane}
\affiliation{Theoretical Division, Los Alamos National Laboratory, Los Alamos, New Mexico 87545, USA}

\author{Jian-Xin Zhu}
\affiliation{Center for Integrated Nanotechnologies, Los Alamos National Laboratory, Los Alamos, 87545, New Mexico, USA }               
\affiliation{Theoretical Division, Los Alamos National Laboratory, Los Alamos, New Mexico 87545, USA}

\date{\today} 
\begin{abstract}
We investigate the electronic and magnetic
properties of copper-doped Pb$_{10}$(PO$_4$)$_6$O using a combination of density functional theory and many-body perturbation theory. The flat half-filled electronic band at the Fermi level is found to give way to an incommensurate antiferromagnetic instability with wave vector $(0.28\pi ,\pm 0.47\pi ,\pi )$ within the random phase approximation arising predominantly from Cu-$d_{yz}$ and Cu-$d_{xz}$ orbitals. Moreover, the Heisenberg exchange coupling between neighboring copper atoms is estimated to be $\sim 1$ meV. Our results suggest that magnetism in copper-doped Pb$_{10}$(PO$_4$)$_6$O is localized on the impurity copper site, with no long-range ordering. These findings support the picture that copper behaves as a magnetic impurity within the Pb-apatite matrix.
\end{abstract}


\maketitle

% Main text
\section{Introduction}\label{sec:introduction}
Flat band systems have recently garnered significant attention as a platform for realizing emergent quantum phases of matter. The negligible band dispersion in these systems yields a nearly infinite effective mass and a large density of states (DOS) near the Fermi level. Consequently, strong correlation effects are readily amplified to drive emergent phenomena, with experimental reports including Fractional Chern insulators\cite{xie2021fractional}, fractional quantum anomalous Hall~\cite{cai2023signatures,park2023observation}, unconventional superconductivity\cite{cao2018unconventional,park2021tunable,tian2023evidence}, strange metals\cite{ye2024hopping}, and non-Fermi liquid behavior\cite{huang2024non}.

Magnetism also naturally arises from flat band dispersions owing to the large DOS which promotes spontaneous spin polarization and lowers the total energy of the system. However, very few material examples have been realized that host flat band driven states. At present, examples are limited to the Kagome family of compounds~\cite{yin2019negative,lin2018flatbands,yin2018giant}, with Lieb lattice predictions suggesting flat band boosted ambient temperature ferromagnetism in the 2D limit\cite{bouzerar2023flat}. These few instances give us a glimpse into the vast phase space of possibilities and call for further exploration of new materials.

Recently, Lee et al.~\cite{lee2023superconductor,lee2023firs} reported a possible room-temperature and atmospheric-pressure superconductor in a copper-doped lead apatite crystal, Pb$_{10-x}$Cu$_{x}$(PO$_{4}$)$_{6}$O for $0.9<x<1.1$, known as LK-99. This announcement attracted great attention that spurred numerous experimental and theoretical groups worldwide to attempt to replicate this extraordinary claims~\cite{kumar2023synthesis, liu2023semiconducting, timokhin2023synthesis, thakur2023synthesis, singh2023experimental, guo2023ferromagnetic, wu2023successful, jain2023superionic, zhu2023first, habamahoro2024replication, kumar2023absence,cabezas2023theoretical, oh2023s, tavakol2023minimal, griffin2023origin, kurleto2023pb, mao2023wannier, bai2023semiconductivity, shen2023phase, paudyal2024implications, liu2023different, karki2024emergence}. The experimental efforts have faced significant challenges in reproducing the synthesis and observations of Lee et al., with a few studies still on-going\cite{wang2024possible,wang2024observation}.

In parallel, the electronic structure of LK-99 was explored in a number of theoretical studies~\cite{cabezas2023theoretical, oh2023s, tavakol2023minimal, griffin2023origin, kurleto2023pb, mao2023wannier, bai2023semiconductivity, shen2023phase, paudyal2024implications, liu2023different, karki2024emergence}. Interestingly, replacing one Pb atom with that of copper transforms the electronic states of lead apatite, yielding a very narrow half-filled band at the Fermi level\cite{jiang2023pb}, thereby hinting at the potential for emergent strong correlation effects. By applying density functional theory (DFT) calculations with a Hubbard $U$, one found ferromagnetic (FM) and A-type antiferromagnetic (AFM-A) states~\cite{bai2023semiconductivity}. Additionally, when spin-orbit coupling (SOC) is included, the flat band splits revealing Weyl nodes at the Fermi level, suggesting LK-99 to be a Weyl semimetal\cite{brass2024weyl,hirschmann2024symmetry}. Beyond DFT calculations have focused on examining correlation corrections to the electronic band structure and possible pairing scenarios\cite{witt2023no,korotin2023electronic,yue2023correlated,kim2024non,sun2023metallization,celiberti2023pb,pashov2023multiple,shimizu2023magnetic}. While the condensed matter community has failed to confirm any of the alleged superconducting properties, the very narrow flat bands at the Fermi level are intriguing and motivate further theoretical study.

In this work, we investigate the electronic and magnetic properties of Pb$_9$CuP$_6$O$_{25}$ using a combination of density functional theory (DFT) and many-body perturbation theory. Our analysis reveals that Pb$_9$CuP$_6$O$_{25}$ is metallic with strong hybridization of Cu-$3d$ and O-$2p$ states near the Fermi level and an incommensurate magnetic instability with a propagation vector of $(0.28\pi, \pm0.47\pi,\pi)$. Furthermore, we estimate the nearest-neighbor Heisenberg exchange coupling parameters to be quite small, $\sim 1$ meV. Thus, despite the seductive nature of the flat band, our results suggest that magnetism in Pb$_9$CuP$_6$O$_{25}$ is localized on the impurity copper site, with no long-range ordering. These findings underscore the picture that copper behaves as a magnetic impurity within the Pb-apatite matrix.

\section{Theoretical and Calculation Details}
\subsection{First-principles calculations}

Ab initio calculations were carried out by using the pseudopotential projector-augmented wave method\cite{blochl1994projector, kresse1999ultrasoft} implemented in the Vienna {\it ab initio} simulation package (VASP)\cite{kresse1996efficient,kresse1993ab} with an energy cutoff of 520 eV for the plane-wave basis set. Exchange-correlation effects were treated using the recently constructed strongly-constrained and appropriately-normed (SCAN) meta-GGA density functional~\cite{sun2015strongly, sun2016accurate}. A $10\times10\times10$ $\Gamma$-centered $k$ mesh was used to sample the Brillouin zone. We substituted a Cu atom for Pb(1) in the pristine crystal structure \cite{krivovichev2003crystal,lai2024first} in accord with the experimental measurements to initialize our calculations. All atomic sites in the unit cell along with the unit cell volume and shape were relaxed simultaneously using a conjugate gradient algorithm to minimize energy with an atomic force tolerance of 0.001 eV/\AA~and a total energy tolerance of $10^{-6}$ eV. The resulting lattice parameters $a$, $b$, and $c$ are equal to $9.643$~\AA, $9.643$~\AA, and $7.233$~\AA, respectively.

\subsection{Local Projections and Response Function Calculations}
The many-body theory calculations of the spin–orbital fluctuations and magnetic instabilities within the random phase approximation (RPA) were performed using real-space local projections, as implemented in VASP\cite{schuler2018charge}. For Pb$_{9}$Cu(PO$_{4}$)$_{6}$O, the full manifold of Cu-$3d$, and the O-$p$ states of the four oxygens with the largest contribution to the electronic density of states around Fermi energy (a total of 17 orbitals) were included in generating the orbital projections.

To calculate the polarizability $\chi_{0}(\mathbf{q},\omega)$, we assume a non-interacting ground state, which allows us to replace the
dressed single particle Green’s function with its non-interacting counterpart,
\begin{equation}
\begin{aligned}
\hat{g}_{\text{KS}}(\mathbf{k},i\omega)&=\sum_{n} \frac{\ket{\psi_{\mathbf{k}n}^{KS}} \bra{\psi_{\mathbf{k}n}^{KS}}}{i\omega - \epsilon_{\mathbf{k}n}}
\end{aligned}
\end{equation}
given in the Kohn-Sham basis where $n$ is the band index and $\mathbf{k}$ is the momentum wave vector in the Brillouin zone. To obtain the atomic-orbital resolved single-particle Green's function the Kohn-Sham wave function is projected onto a set of localized atomic-site-, orbital- and spin-dependent basis functions $\ket{\phi^{\tau\sigma}_{lm}}$\cite{pizzi2020wannier90}:
\begin{equation}
\ket{\psi_{\mathbf{k}n}} =\sum_{lm\tau} P^{lm\tau}_{nk\sigma} \ket{\phi^{\tau\sigma}_{lm}}
\end{equation}
where $P^{lm\tau}_{nk\sigma}$ is the overlap between the Kohn-Sham wave function and the local basis set, $\braket{ \phi^{\tau\sigma}_{lm}|\psi_{\mathbf{k}n} }$. $\hat{g}_{\text{KS}}(\mathbf{k},i\omega)$ is then projected onto the local basis set as:
\begin{equation}
\begin{aligned}
G^{\sigma\sigma'}_{0~lm\tau,l'm'\tau'}(\mathbf{k}, i\omega) &\equiv \bra{\phi^{\tau\sigma}_{lm}}\hat{g}_{\text{KS}}(\mathbf{k},i\omega)\ket{\phi^{\tau'\sigma'}_{l'm'}} \\
&=\sum_{n} \frac{P^{lm\tau}_{nk\sigma} P^{*l'm'\tau'}_{nk\sigma}}{i\omega - \epsilon_{nk}} 
\end{aligned}
\end{equation}
This Green's function is analogous to that obtained using a tight-binding Hamiltonian, where the eigenvectors of the Hamiltonian have been replaced by the local projector overlaps. Furthermore, this form allows for the computation of the non-interacting polarizability:
\begin{equation}
\begin{aligned}\label{eq:Pol}
&{\chi_{0}}^{\sigma\sigma'\sigma''\sigma'''}_{\mu\nu\mu'\nu'}(\mathbf{q}, i\omega_n) = \\
&\frac{1}{N_{\mathbf{k}}} \frac{1}{\beta}\sum_{\mathbf{k}l} G^{\sigma\sigma'}_{0~\mu\nu}(\mathbf{q+k}, i\omega_n + ip_l) G^{\sigma''\sigma'''}_{0~\mu'\nu'}(\mathbf{k}, ip_l)
\end{aligned}
\end{equation}
where $i\omega_n$ and $ip_l$ are Matsubara frequencies, $\beta=1/k_BT$, $T$ is the temperature and $k_B$ is Boltzmann constant, and $\mu(\nu)$ are composite indices of orbital and atomic site. Using $G_0$, the polarizability can be simplified by performing the Matsubara frequency summation and analytically continuing $i\omega_n \rightarrow \omega+i\delta$, for $\delta\rightarrow 0^{+}$. Implementation details of Eq.~(\ref{eq:Pol}) are given in Ref.~\cite{lane2023competing}.

\begin{figure*}[ht]
    \centering
    \includegraphics[width=0.99\textwidth]{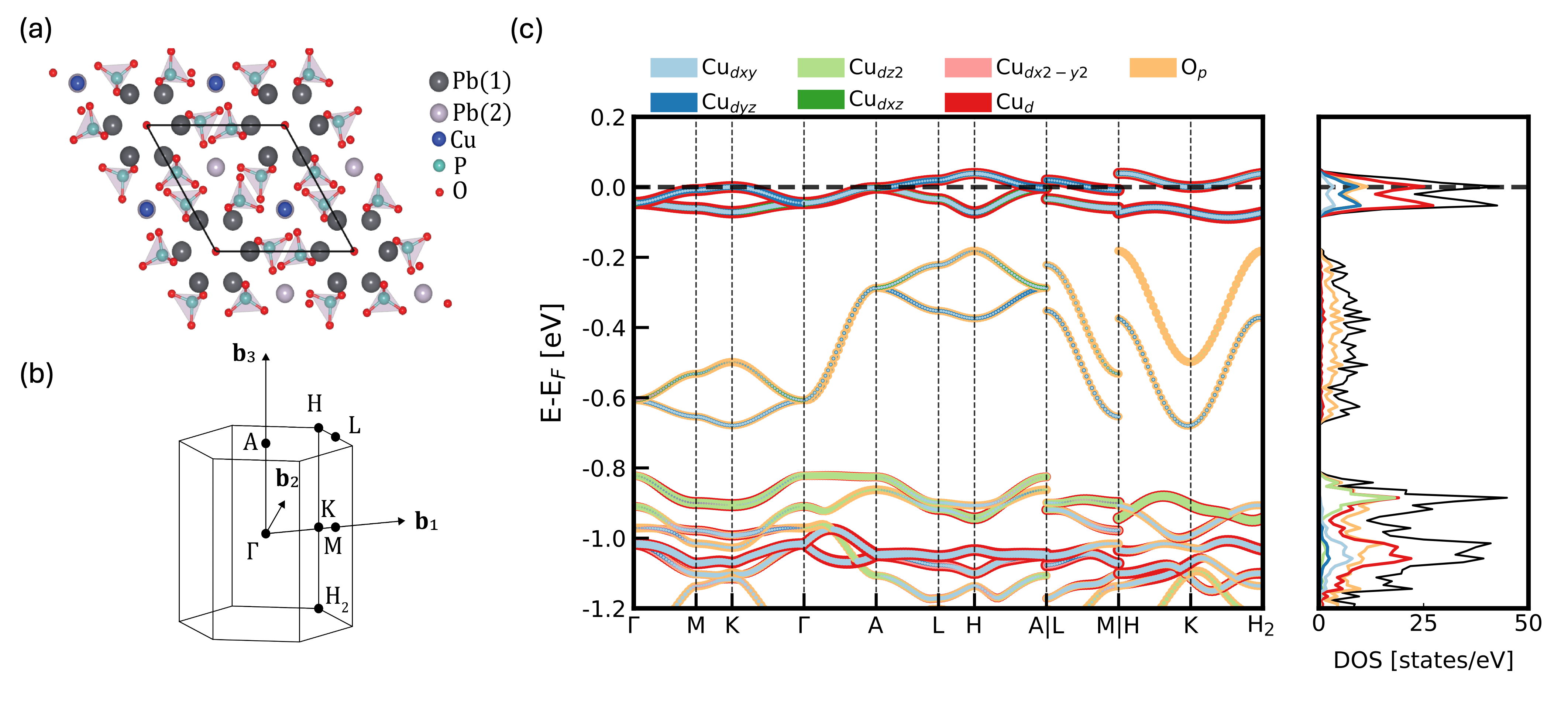}
    \caption{(a) Crystal structure of Pb$_9$CuP$_6$O$_{25}$ viewed along the $c$-axis. Black lines denote the primitive unit cell. (b) The associated first Brillouin zone with labeled high-symmetry points. (c) Electronic band structure and corresponding density of states for the non-magnetic phase. The size and color of the dots are proportional to the fractional weight of the various indicated site-resolved orbital projections.}
    \label{fig:band}
\end{figure*}

To gain insight into the landscape of charge and magnetic instabilities in the ground state, we examine the response ($\delta \rho$) of the system to an infinitesimal perturbing source field ($\delta \pi$). The associated response function in the generalized RPA-type matrix form is given by,
\begin{equation}\label{eq:chirpa}
\chi^{MN}(\mathbf{q}, \omega) = \left[1 - \chi_0^{MI}(\mathbf{q}, \omega) v^{IK}\right]^{-1} \chi_0^{KN}(\mathbf{q}, \omega)
\end{equation}
where the orbital indices have been suppressed, the spin indices I,J,L,M take the value $0,x,y,$ or $z$, $v^{ML}$ is the electron–electron interaction, and the polarizability $\chi_0$ is defined in Eq.~\ref{eq:Pol}. To facilitate our analysis, we let 
\begin{align}\label{eq:FandFeig}
   \bar{F}^{MK}(\mathbf{q},\omega) &=  \chi_0^{MI}(\mathbf{q}, \omega) v^{IK}\\
   &=V_\alpha(\mathbf{q}, \omega) \Lambda^\alpha_F(\mathbf{q}, \omega) V^{-1}_\alpha(\mathbf{q}, \omega)\label{eq:FFeig}
\end{align}
where \(\Lambda_F\) is a diagonal matrix and \(V\) is the eigenvectors. Now as \(\Lambda_F(\mathbf{q},\omega=0)\) approaches $1$, $\chi$ becomes singular, signaling an instability to an ordered phase with ordering vector $\mathbf{Q}$. 

A $21 \times 21 \times 21$ $\Gamma$-centered mesh was used to evaluate the local projectors and the response functions in the Brillouin zone. Only the 8 bands with in a $1$ eV window about the Fermi level were retained in the band summation in the polarizability. A small broadening $\delta=0.01$ eV was used with a temperature of $\text{T}=10 \mathrm{~K}$. Coulomb interactions were included via the multi-orbital Hubbard model on the Cu sites similar to Ref.~\cite{lane2022identifying}. The maximum instability was found to equal $1$ for $U=0.67$ eV.

\section{Results and discussion}
\subsection{Crystal and Electronic Structure}
Figure~\ref{fig:band}(a) and (b) display the crystal structure in the primitive cell and the first Brillouin zone with the high-symmetry points labeled for Pb$_{9}$Cu(PO$_{4}$)$_{6}$O, respectively. The lattice is made up of layered planes containing Cu, O, P and Pb atoms. In this structure, Cu atoms form one-dimensional chains along the c-axis, connecting through oxygen atoms in a linear arrangement (Cu-O-Cu). Notably, only a single copper atom appears in the unit cell. P and O atoms form phosphate (PO$_4$) tetrahedrons, which help stabilize the lattice, while Pb atoms occupy sites in a layered arrangement. This structure is built on a trigonal Bravais lattice with P3 symmetry and is non-centrosymmetric. 

Figure \ref{fig:band}(c) presents the electronic band structure of Pb$_9$CuP$_6$O$_{25}$ in the non-magnetic (NM) phase. Strong hybridization between the Cu-$3d$ and O-$2p$ orbitals is observed, where the half-filled antibonding Cu-$3d$/O-$2p$ band crosses the Fermi level and its bonding partner ``bookends'' the spectrum from 1.1 eV binding energies. In between, the dispersive electronic states are primarily contributed by O-$2p$ orbitals. The Cu-$3d$ bands are very narrow, especially near the Fermi level. In contrast, the O-$2p$ bands are about twice more dispersive than the Cu-$3d$ bands in the $b_{1}$-$b_{2}$ plane, and display strong dispersion along the $b_{3}$ axis, as illustrated along the $\Gamma$-A, L-M, H-K, and K-$H_{2}$ lines in the Brillouin zone. This suggests the Cu-$3d$ electrons are quite localized in the unit cell, whereas the O-$2p$ electrons can readily hop between the Pb layers along the $c$-axis. The flat bands at the Fermi level arise from the partial (1/4) occupancy of the corner oxygen sites and the hybridization with the Cu-$3d$ orbitals~\cite{tavakol2023minimal,griffin2023origin,kurleto2023pb}. 

Typically, when a compound is doped via chemical substitution we witness the appearance of donor/acceptor states as in semiconductors or the addition/subtraction of electrons (holes) from a metallic system resulting in a shift of the Fermi level. Overall, the nature of the electronic states is expected to be adiabatically connected to the parent undoped compound. Here, when a single lead atom is substituted by copper atom, the electronic structure undergoes a dramatic change from an insulator to a flat band metal. This suggests two pictures, either copper is an impurity that is highly hybridized with the rest of the Pb-apatite system or Pb$_9$CuP$_6$O$_{25}$ is the parent compound of a new family of flat band materials with little connection to its insulator ancestor.

\subsection{Magnetic Instabilities}
A key consequence of flat bands at the Fermi level is the possible emergence of broken symmetry phases that have the capacity to couple to multiple different degrees of freedom. In Pb$_9$CuP$_6$O$_{25}$, reports of magnetism~\cite{guo2023ferromagnetic,puphal2023single,liu2023phases,wang2023ferromagnetic} are intriguing and are in need for further analysis. 

\begin{figure*}[ht]
	\centering
	\includegraphics[width=0.99\textwidth]{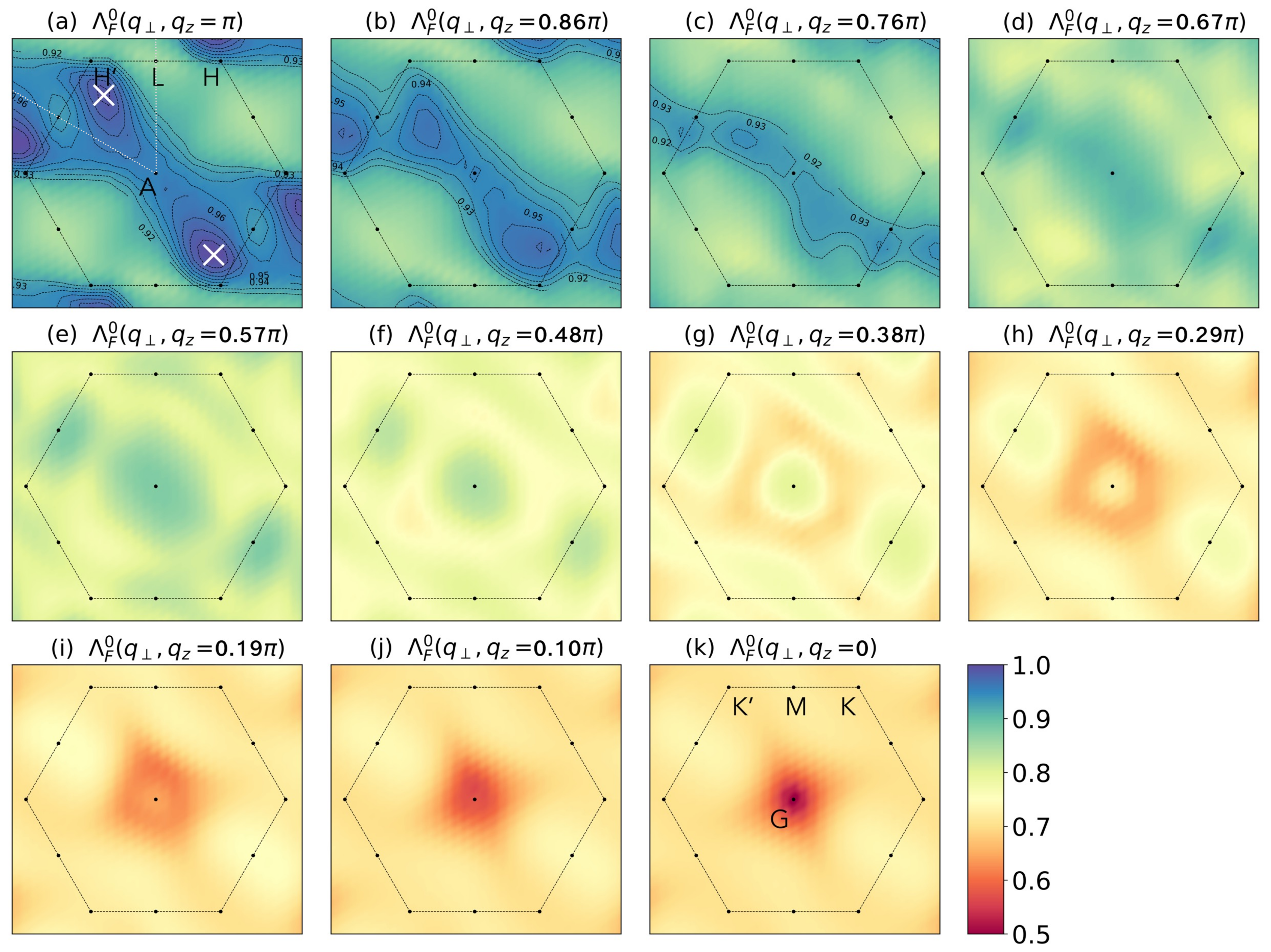}
    \caption{Momentum dependence of $\Lambda_F^0(\mathbf{q}, \omega=0)$ for pristine Pb$_9$CuP$_6$O$_{25}$ in the non-magnetic phase for various slices along $q_z$. The white “$\times$” marks denote the critical instability momenta $\mathbf{Q}$. The black dashed line gives the boundary of the Brillouin zone and the gray dotted lines indicate the reciprocal lattice vectors. The color bar indicates the instability strength. 
    }
	\label{fig:EigF}
\end{figure*}

Figure~\ref{fig:EigF} shows the leading instbility $\Lambda_F^0(\bold{q},0)$ for various values of $q_z$ between $\pi$ and 0 for Pb$_9$CuP$_6$O$_{25}$.
In panel (a), $\Lambda_F^0(\mathbf{q}_\perp,0)$ reaches a value of $1$ at $\mathbf{Q} = (0.28\pi, \pm0.47\pi,\pi)$ (white `$\times$'). The ordering vectors are incommensurate and lie off of the high-symmetry lines of the Brillouin zone. Moreover, $\Lambda_F^0$ is very flat throughout the zone, as indicated by the overlaid contour curves. This flatness suggests that there is a dense manifold of $q$-vectors that compete with $\mathbf{Q}$ and hint at the presence of magnetic fluctuations in the ground state of Pb$_9$CuP$_6$O$_{25}$. 

As $q_z$ decreases, the maximum value of $\Lambda_F^0$ is reduced with a concomitant contraction of the plateau of instabilities.  In Panels (b)-(f), there are three `islands' around the $\Gamma$ and $M$ points in the Brillouin zone (green shaded regions), whereas in Panels (g)-(k) the instability plateau at the zone center reduces significantly, forming a ring-like structure.  Finally, the minimum value of $\Lambda_F^0$ is concentrated about $\Gamma$, thereby indicating that ferromagnetic (FM) ordering is the least favorable magnetic configuration.

In single crystal samples, no anomalies indicative of phase transitions are observed over a wide range of temperatures up to $800$ K, consistent with a fluctuating magnetic state\cite{puphal2023single}. Since the flat band is extremely sensitive to structural and substitutional disorder, some experiments\cite{guo2023ferromagnetic,puphal2023single,liu2023phases,wang2023ferromagnetic} observe a weak ferromagnetism. However, this is likely due to the local magnetic order in inhomogeneous networks of Cu substituents that originates from frustrated exchange interactions\cite{puphal2023single}. 

To quantify the total number of competing magnetic configurations, we introduce the density of instabilities $\lambda(\Omega)$ similar to Ref.~\cite{lane2023competing}, where $\lambda(\Omega)d\Omega$ is the number of instabilities in the system whose strengths lie in the range from $\Omega$ to $\Omega+\delta\Omega$. That is, $\lambda(\Omega)$ is defined as
\begin{align}
\lambda(\Omega)=\frac{1}{N_\mathbf{q}}\sum_{\mathbf{q}\alpha}\delta\left( \Lambda_F^\alpha(\bold{q},0)-\Omega\right)
\end{align}
where $\Omega$ is the instability strength and $\alpha$ enumerates the instability eigenvalues defined in Eq.~\ref{eq:FFeig}. We further emphasize that $\lambda(\Omega)$ contains the instability information for all eigenvalues, not just for the maximum.

\begin{figure}[t]
	\centering
    \includegraphics[width=1\linewidth]{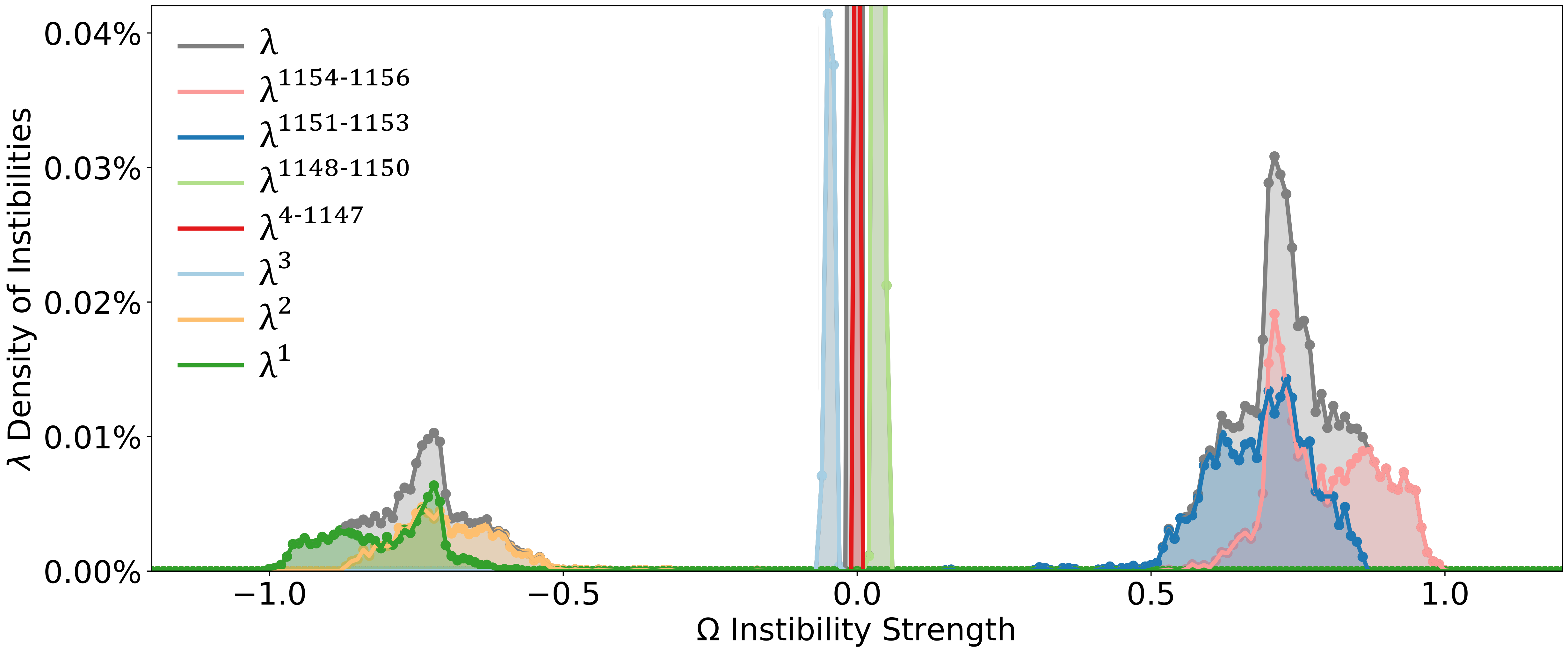}
	\caption{Density of instabilities for pristine Pb$_9$CuP$_6$O$_{25}$ in the non-magnetic phase. Shading and lines of various colors (see legend) give the contributions from the total and different instability eigenvalues. }
	\label{fig:DosF}
\end{figure}

Figure~\ref{fig:DosF} presents the density of instabilities $\lambda(\Omega)$ for Pb$_9$CuP$_6$O$_{25}$, where the colors indicate the contribution of various instability eigenvalues. The total density of instabilities (grey line and shading) displays  sharp leading edge near an instability strength of $1.0$ arising from the flat region of $\Lambda_F^0$ near $Q$ [Fig.~\ref{fig:EigF} (a)-(c)]. The large population of instabilities at the leading edge with different ordering $\mathbf{q}$-vectors suggests competition between $\mathbf{q}$-vectors that typically necessitates the inclusion of vertex corrections to capture the low temperature behavior\cite{moriya2012spin}.

A pronounced Van Hove peak appears around $\Omega\sim0.7$ that originates from the change in concavity of the ring-like feature around $\Gamma$ from negative to positive  [Fig.~\ref{fig:EigF} (d)-(h)]. Finally, $\lambda(\Omega)$ exhibits a tail for $0.3<\Omega<0.55$ driven by the localized minima at $\Gamma$ [Fig.~\ref{fig:EigF} (k)]. Overall, $\lambda(\Omega)$ displays a distribution similar to that of 2D system, wherein a Van Hove singularity is bookended by two step edges\cite{van1953occurrence}, despite the 3D bonding network of the crystal. The effective dimensionality of the fluctuations may be driven, in part, by the anisotropy in electron hopping between in- and out-of-plane, where electrons appear more localize in-plane and delocalized along the $c$-axis. 

Interestingly, $\lambda(\Omega)$ is symmetric about $\Omega=0$, albeit the spectrum for $\Omega<0$ is reduced by a factor of three. To understand this, we break $\lambda(\Omega)$ down into its instability eigenvalues. We find seven distinct contributions [colored lines and shading in Fig.~\ref{fig:DosF}], three above (below) zero and a large peak at $\Omega=0$. Those above (below) zero are triply (non-)degenerate. To explain these degeneracies, we consider the multi-orbital Hubbard model for $v$, which yields \(v^{00} = U/2\) and \(v^{xx,yy,zz} = -U/2\), and the susceptibility matrix has the form \(\chi^{00} =\chi^{xx}=\chi^{yy}=\chi^{zz}= \chi_0\) in the absence of spin-orbit coupling and spin-polarization. Inserting these into Eq.~\ref{eq:FandFeig}, 
\begin{equation}
\bar{F}(\mathbf{q},0) =\frac{U}{2} \text{diag}\left(
\chi_0 , -\chi_0 , -\chi_0 , -\chi_0\right),
\label{eq:F_matrix}
\end{equation}
therefore, the eigenvalues of \(\bar{F}\) are straightforwardly given by:
\begin{equation}
\Lambda_F(\mathbf{q},0) = \frac{U}{2} \text{eig}(\chi_0)\left\{1, -1, -1, -1\right\}.
\end{equation}
This means Fig.~\ref{fig:DosF} the distribution of the various colored band contributions stem from the multi-orbital nature of $\chi$, whereas the degeneracies and symmetry about $\Omega=0$ arise from spin being a good quantum number.

\begin{figure}[hb]
    \centering
    \includegraphics[width=1\linewidth]{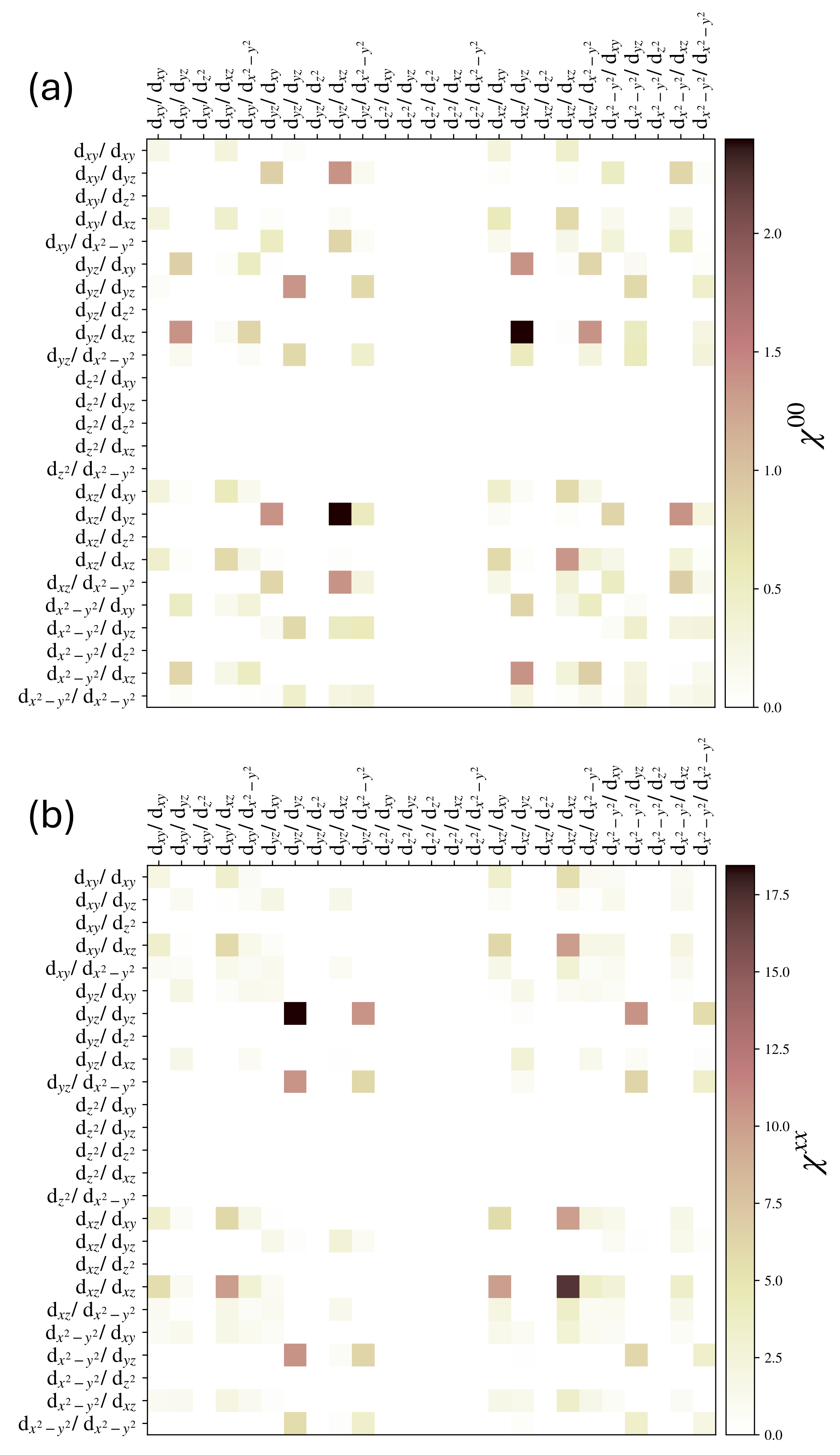}
    \caption{Real part of the RPA susceptibility $\chi$ for various combinations of Cu $d$ orbitals at upper left Q point white in Fig\ref{fig:EigF}: (a) $\chi^{00}$ charge channel and (b) $\chi^{xx}$ spin channel.}
    \label{fig:chi_rpa}
\end{figure}

Figure~\ref{fig:chi_rpa} presents a heat map of the various atomic-site and orbital matrix elements of (a) $\chi^{00}$ (charge-channel) and (b) $\chi^{xx}$ (spin-channel) evaluated at $\mathbf{Q}$. Since $\chi^{xx} = \chi^{yy} = \chi^{zz}$, we focus our discussion on $\chi^{xx}$. The charge channel susceptibility $\chi^{00}$ [Fig.~\ref{fig:chi_rpa} (a)] displays maxima along the off-diagonal components that involve $(d_{xz}/d_{yz},d_{yz}/d_{xz})$ and $(d_{yz}/d_{xz},d_{xz}/d_{yz})$ orbital combinations. This is consistent with the orbital character of the electronic band structure near the Fermi level [Fig.~\ref{fig:band}(c)]. 

The spin-susceptibility is an order of magnitude larger than the charge susceptibility, where diagonal matrix elements dominate. Specifically, $\chi^{xx}$ is primarily comprised of copper $(d_{yz}/d_{xy},d_{yz}/d_{xy})$, and $(d_{xz}/d_{xz},d_{xz}/d_{xz})$ orbital contributions. Additionally, several off-diagonal terms exhibit notable contributions, such as $(d_{xz}/d_{xz},d_{xy}/d_{xz})$ and $(d_{yz}/d_{x^2-y^2},d_{yz}/d_{yz})$, and their equivalents. Other terms, such as $(d_{xy}/d_{xz},d_{xy}/d_{xz})$ and $(d_{xy}/d_{xz},d_{xz}/d_{xy})$, and their equivalents, weakly contribute. Interestingly, the Cu-$d_{z^2}$ orbital minimally contributes to $\chi^{xx}$, since it primarily originates from bands far from the Fermi level. 

In both spin and charge channels, $O$-$2p$ play a very limited role in the charge/spin fluctuations compared to those from copper. This stems from their very small weight in the electronic structure near the Fermi level and the negligible correlations on the O sites. Moreover, the reduced weight in the charge channel is driven, in part, by the use of the multi-orbital Hubbard model for $v$. That is, the local nature of $v$ promotes local moment formation by setting an energy penalty for two spin to occupy the same site. In contrast, purely local interactions cannot encourage charge disproportionation since they do not provide any energetic advantage for accumulating (depleting) charge on neighboring atomic sites. 

\subsection{Magnetic Exchange Parameters}
The fluctuating magnetic state described in the RPA analysis can originate from competition between different long-range magnetic configurations or a simple local impurity, so to distinguish these two scenarios, we estimate the nearest-neighbor exchange coupling strength between copper atoms. To accomplish this, we map the total energies of the various spin configurations $\gamma$ onto those of the nearest-neighbor Heisenberg Hamiltonian~\cite{xiang2013magnetic,zhu2010band},  written as
\begin{equation}
\mathcal{H}=\sum_{\braket{i<j}}J_{ij}\mathbf{S}_{i}\cdot \mathbf{S}_{j}.
\label{eq:Hnn}
\end{equation}
Here $i(j)$ indexes the magnetic ion lattice positions, $S_{i}$ is the local magnetic moment on lattice site $i$. $J_{ij}$ is the nearest-neighbor exchange interaction strength, with the symmetry properties of $J_{ij}=J_{ji}$. Assuming a fully isotropic bond independent model and separating in- and out-of-plane components, we find 
\begin{align}
J_\perp=\frac{(E^{\uparrow\uparrow}+E^{\downarrow\downarrow})-(E^{\uparrow\downarrow}+E^{\downarrow\uparrow})}{4\cdot 4\braket{S}^{2}},\\
J_\parallel=\frac{(E^{\uparrow\uparrow}+E^{\downarrow\downarrow})-(E^{\uparrow\downarrow}+E^{\downarrow\uparrow})}{4\cdot 2\braket{S}^{2}}.
\end{align}
Using our SCAN ground state total energies, \( J_{\perp} = 0.47 \ \text{meV} \), while \( J_{\parallel} = -1.67 \ \text{meV} \), with a self-consistent copper magnetic moment of $0.73\mu_B$. The magnetic exchange values are quite small, consistent with the large in-plane and out-of-plane Cu-Cu distance of 9.66 \(\text{\AA}\) and 7.23 \(\text{\AA}\), respectively. Additionally, the neighboring oxygen atoms exhibit a small but non-zero magnetic moment of \(\sim \pm 0.03 \mu_B\), while the magnetic moments of Pb and P atoms are effectively zero (\(\sim \pm 0.005 \mu_B\)), suggesting the oxygens form a weak bridge to help facilitate the exchange interactions between Cu atoms. 

The dichotomy of localized and extended states finds an intersection in flat band systems. Flat bands may arise from localized impurity levels, low dimensional sub-units of solids, or geometric frustration in lattices, but the physical underpinnings are quite different. Despite the appearance of a flat band at the Fermi level in copper-doped Pb-apatite, the band character is heavily copper, suggesting these states are trivial impurity-like states. Furthermore, in comparing to our RPA and magnetic exchange coupling analysis, the weak coupling parameters indicate the magnetic moments are quite local with little connection to their neighboring magnetic ion sites, allowing for multiple magnetic ground state that are effectively degenerate. This suggests the RPA results should be interpreted as stemming from local copper atoms that are isolated and do not significantly couple with its environment, thereby generating a multiplicity of seemingly degenerate ground states. To corroborate this picture we attempted to calculate the next-nearest and next-next-nearest exchange parameters, but we could not accurately converge our total energies below $\sim0.1$ meV scale, thereby giving more credence to localized impurity moment picture.

\section{Conclusion}
In summary, we have performed density functional theory and RPA calculations of the electronic band structure, magnetic instabilities, and magnetic exchange parameters of copper-doped Pb\textsubscript{10}P\textsubscript{6}O\textsubscript{25}. We reveal an extremely flat half-filled band at the Fermi level composed of hybridized Cu-$d$ and O-$p$ states where upon adding on-site Coulomb interactions, a fluctuating incommensurate AFM instability is predicted arising from Cu-$d_{yz}$ and -$d_{xz}$ orbitals. However, despite the alluring nature of the flat band, the weak nearest-neighbor exchange coupling combined with the RPA analysis, suggest the magnetism in Pb\textsubscript{9}CuP\textsubscript{6}O\textsubscript{25} is localized on the impurity copper site.  

\section*{Declaration of competing interest}
The authors declare that they have no known competing financial interests or personal relationships that could have appeared to influence the work reported in this paper.

\section*{Acknowledgments}
This work was carried out under the auspices of the U.S. Department of Energy (DOE) National Nuclear Security Administration under Contract No. 89233218CNA000001. It was supported by the LANL LDRD Program, and in part by the Center for Integrated Nanotechnologies, a DOE BES user facility, in partnership with the LANL Institutional Computing Program for computational resources. Additional computations were performed at the National Energy Research Scientific Computing Center (NERSC), a U.S. Department of Energy Office of Science User Facility located at Lawrence Berkeley National Laboratory, operated under Contract No. DE-AC02-05CH11231 using NERSC award ERCAP0028014.  

\bibliographystyle{unsrtnat}

\bibliography{LK99}

\end{document}